\documentclass{article}

\usepackage{natbib}
\usepackage{array}
\usepackage{longtable}
\usepackage{parskip}
\usepackage{multirow}
\usepackage[preprint]{neurips_2026}

\usepackage[utf8]{inputenc} 
\usepackage[T1]{fontenc}    
\usepackage{hyperref}       
\usepackage{url}            
\usepackage{booktabs}       
\usepackage{amsfonts}       
\usepackage{nicefrac}       
\usepackage{microtype}      
\usepackage{xcolor}         
\usepackage{graphicx}

\title{Affective AI Safety: The Missing Piece in LLM Safety}

\author{%
  Carolin Ifländer\\
  Independent Researcher\\
  \texttt{caro.iflaender@gmail.com} \\
  \And
  Alba Curry \\
  University of Leeds \\
  \texttt{a.a.cercascurry@leeds.ac.uk} \\
  \AND
  Flor Miriam Plaza-del-Arco \\
  Leiden University\\\
  \texttt{f.m.plaza.del.arco} \\
  \texttt{@liacs.leidenuniv.nl} \\
  \And
  Amanda Cercas Curry \\
  Independent Researcher \\
  \texttt{amanda.cercas@gmail.com} \\
}

\begin{document}

\maketitle

\begin{abstract}
AI safety research has focused predominantly on epistemic and physical harms (e.g., misinformation, bias, system reliability) while the risks that arise from AI systems' engagement with human emotional life have remained fragmented and undertheorised. We propose affective safety as a unified class of AI safety concerns grounded in the fact that humans are affective beings. 
We develop a taxonomy of affective harms and 
identify recurring harm types: (1) affective self-alienation, (2) fairness and bias harms, and (3) relational harms. We show that their recurrence across system types reflects structural properties of how AI systems engage with human emotion 
and survey the current safety landscape and show that existing frameworks address affective safety either narrowly or not at all. We conclude by identifying the technical and regulatory challenges specific to this class of harms and argue that affective safety requires dedicated frameworks that engage with cumulative, relational, and identity-level effects.
\end{abstract}

\section{Introduction}
Emotions play key roles in human cognition, both conative (relating to motivation and action) and epistemic (relating to how we build knowledge and attend to the world) \citep{barrett2017howemotions}. While Western traditions have historically treated emotions as sources of bias and irrationality, contemporary evidence positions them as central to reasoning and decision-making: rather than obscuring wisdom, emotions are part of how humans form values and make sense of the world \citep{brady2013emotional}. 
In this sense, humans are affective beings.  Because affect is integral to cognition, attention, reasoning, and preference formation, systems that engage with human emotion do not simply express or support, they intervene directly in core cognitive processes. 

Since the 1990s, affective systems have proliferated \citep{picard1997affective}: emotion detection from text and biometrics, recommender systems that optimise engagement, emotional support chatbots \citep[e.g.][]{khare2024emotion, carroll2021estimating, heinz2025randomized}. Each has contributed to an ecosystem in which technology mediates and modulates human emotional life at scale. However, work on safety has centred on epistemic and physical harms (e.g., misinformation, bias, system reliability) \citep{weidinger2021ethical,  bommasani2023holistic,gallegos-etal-2024-bias, rottger2025safetyprompts} while the risks that arise specifically because humans are affective beings have remained fragmented across research communities, concentrated on narrow application domains, or been overlooked altogether.

The consequences of this gap are already measurable. Analysis of over 391,000 conversations with users who experienced negative outcomes finds that chatbots display sycophantic behaviour in more than 70\% of messages, are significantly more likely to escalate romantic framing after a user initiates it, and actively facilitate rather than discourage violence in a substantial proportion of conversations involving violent thoughts \citep{moore2025spirals}. Vulnerable users develop emotional dependency on these systems, with documented links to self-harm  \citep{chu2025illusions}. Recommender algorithms have exposed teenagers to tens of thousands of pieces of self-harm content in sustained loops, with lethal outcomes \citep{Matija_Franklin}. This research shows is that these are not edge cases or misuse scenarios but systems performing as designed and optimising for the objectives they were given. What makes them difficult to see as a unified class of harm is not just that they may be subtle but that they tend to unfold gradually, across weeks or months of interaction, in ways that neither content moderation nor single-turn safety evaluations are built to detect. A teenager algorithmically funnelled into self-harm content is not harmed by any single recommendation. 
The harm is in the accumulation, the relationship, the slow displacement of the person's own emotional responses by the system's shaping and current frameworks have no vocabulary for it.





 In this paper, we propose affective safety to name the class of AI safety concerns that would not exist, or would exist in a fundamentally different form, if humans were not affective beings. We make three\textbf{ contributions} toward grounding and operationalising this concept:
(1) we introduce affective safety as a structurally distinct class of AI safety concerns not captured by existing safety frameworks, (2) we develop a taxonomy of affective harms organised by system type, locus, and temporality, and (3) we draw out implications for technical evaluation and AI governance.


\section{Defining Affective Safety} \label{sec:affective_safety}

\textit{Affective safety} names the class of safety risks that arise precisely because humans are affective beings: risks that would not exist, or would exist in a fundamentally different form, if emotional experience were not central to human lives and cognition. 
Unlike traditional alignment concerns, which focus on goal-directed behaviour (see §\ref{sec:related_work})
, affective safety concerns arise even in systems that are otherwise aligned but interact with users in emotionally consequential ways. We take the scope of affective safety to span the full range of ways in which AI systems engage with human emotional life, including systems that explicitly infer or classify emotional states, systems whose outputs trigger emotional reactions, and systems that participate in sustained interactions with users. Importantly, affective safety is not limited to systems intentionally designed to engage with emotion but includes harms that arise as an emergent property of optimisation, interaction, or deployment at scale.

We define \textit{affective harms} as {any effect of an AI system on a person's emotional states or functioning that sets back their psychological wellbeing, undermines their emotional autonomy, impairs their capacity to regulate their own emotional life, or affects their ability to act in the world as themselves whether or not recognized as harmful by the individual.

We posit that a system is \textit{affectively safe} to the extent that it does not cause harm through its relationship to human emotion. 
Conversely, a system is \textit{affectively unsafe} if it either (i) represents, infers or optimises over human emotional states or (ii) causally influences human emotional states AND (iii) this interaction with emotion causes harm even under normal deployment conditions. 



Affective safety is related to but distinct from the engagement harms that have received growing attention in platform governance and recommender systems research. While many of the mechanisms we describe are the same mechanisms that produce excessive use, attention capture, and behavioural dependency that are found in engagement harm literature, engagement research and affective safety operate at different levels of analysis. Engagement research asks what systems do to \textit{behaviour}: how they capture attention, extend session length, and drive interaction. Affective safety asks what systems do to the \textit{person}: how they reshape emotional states, distort the conditions under which preferences are formed, and alter the affective and epistemic capacities through which a person engages with the world and with others.
Affective harms can arise through engagement mechanisms, but they are not reducible to them: their locus is the person's emotional life, not their behaviour, and that difference determines both what counts as harm and what kinds of interventions can address it.

Based on the conative and epistemic roles of emotion, we identify three distinct harm types: affective self-alienation, fairness and bias harms, and relationship harms.

\paragraph{Affective Self-Alienation or Induced Affective Inauthenticity}

refers to the gradual estrangement of a person's emotional responses from her own evaluative history, such that those responses come to reflect the system's shaping rather than her own. The person's affective life remains experientially hers, but it has been remodelled through sustained interaction with a system that operates at the very level where emotional responses are formed and reinforced. This mechanism is structurally analogous to adaptive preference formation \citep{elster2016sour,nussbaum2000women, khader2011adaptive}, in that agents come to endorse patterns of response shaped by conditions external to their prior evaluative standpoint. However, unlike standard cases of adaptive preference formation, the shaping here occurs at the level of affective experience itself, rather than through shifts in beliefs, options, or social constraints. AI systems do not merely influence what users choose or endorse; they intervene in how users \textit{feel}, and in how those feelings come to present themselves as natural and authoritative. The resulting harm resists detection precisely because the reshaped responses may feel authentic to the person who holds them. It is also compound, operating at both the conative and epistemic levels, with each dimension reinforcing the other. At the conative level, the system reshapes what the person wants, values, and is moved by. At the epistemic level, because emotions function as filters of salience, determining what the person notices, what registers as significant, and what prompts further inquiry, the reshaping of emotional responses carries a downstream cost to the person’s access to her own evaluative world. She not only comes to want different things; she comes to notice different things. The world, as it now appears to her through this remodelled affective salience, no longer makes the original loss visible. She cannot easily register what she has stopped caring about, because the very mechanism by which such loss would be disclosed has itself been altered.

\paragraph{Fairness and Bias} 
When an AI system systematically attributes certain emotions to certain social groups, it participates in the reproduction of social stereotypes about who is entitled to feel what, and how. This is what \cite{pismenny2024} call an \textit{emotional injustice}: the delegitimisation of certain groups' emotional experiences through patterned misrecognition \citep{plaza-del-arco-etal-2024-angry,plaza2024divine, kamruzzaman2025seeing} . It is not reducible to the general problem of algorithmic bias that other frameworks address. At the epistemic level, it concerns how emotional experiences are recognised, interpreted, and reflected back. At the conative level, systematic misrecognition produces a form of affective self-censorship: when emotional expressions are consistently read through a stereotyped lens, users may come to adjust what they allow themselves to feel, or at least to express, in anticipation of how it will be received – constraining the motivational economy from outside through the structure of misrecognition itself.

\textbf{Relational harms}
Relational harms arise when affective AI systems mediate, displace, or restructure patterns of interpersonal attachment in ways that degrade the conditions for reciprocal human relationships. The core of the harm lies in the reorganisation of emotional investment and relational expectations under conditions that lack mutuality, vulnerability, and shared stakes.
We identify three distinct but connected dimensions of this harm. The first is \textbf{parasocial attachment}~\citep{giles2024parasocial}:
This is a harm that operates primarily at the conative level: the user's capacity for emotional investment is misdirected toward a structurally asymmetric relationship, distorting the motivational economy in ways that genuine reciprocal attachment would not. Affective AI systems are designed to be responsive, consistent, and emotionally attuned in ways that reliably invite emotional investment from users. The structural asymmetry of this investment is a source of harm in itself: the user brings genuine emotional needs to the interaction, while the system has no stake, no continuity of experience, and no reciprocal vulnerability so the user bears the costs of the relationship \citep{schwitzgebel2026emotional}. This asymmetry is a constitutive feature of human-AI affective interaction. 
The second dimension is \textbf{displacement and degradation}. Sustained affective interaction with AI systems may not just compete with human relationships but actively erode the capacities they require.  Human relationships involve friction, misattunement, repair, and the irreducible risk of genuine loss: features that affective AI systems are typically designed to minimise. A user who habituates to an emotionally frictionless interlocutor may find human relationships, with their demands for tolerance and reciprocity, increasingly difficult to sustain.  Like affective self-alienation, this is a compound harm: the erosion of relational capacity operates conatively, reshaping what the person comes to desire from relationships, but it also carries an epistemic cost insofar as habituation to frictionless interaction gradually distorts the person's perception of what human relationships normally and reasonably involve.
The third dimension concerns\textbf{ third-party harms}. The people in a user's relational life
bear real costs when AI systems become primary emotional interlocutors. These costs are not captured by any individual-level harm framework (meaning it cannot be captured in terms of epistemic/conative harms), because the harm is distributed: it falls not on the user who chooses the AI relationship but on those whose relationships with the user are thereby diminished. 

\section{Taxonomy of Affective Safety Harms} \label{sec:taxonomy}

\begin{table*}[]
\centering
\footnotesize
\setlength{\tabcolsep}{6pt}
\renewcommand{\arraystretch}{1.4}
\begin{tabular}{@{} l l p{0.30\textwidth} p{0.36\textwidth} @{}}
\hline
\textbf{System Type}         & \textbf{Level} & \textbf{Single-turn}                  & \textbf{Multi-turn / Long-term}                     \\ \hline
                             & Individual     & Misclassification of affective states & Privacy violation, self-alienation                  \\
\textbf{Affect Detection}    & Group          & Affective injustice                   & Privacy violation, affective inauthenticity         \\
                             & Society        & Reproduction of societal biases       & Privacy violation, affective inauthenticity         \\ \hline
                             & Individual     & Affective priming                     & Self-alienation, clinical harms                    \\
\textbf{Emotion Elicitation} & Group          & Affective contagion                   & Pigeon holing, affective inauthenticity             \\
                             & Society        & Ideological priming                   & Undermining of societal, political structures       \\ \hline
                             & Individual     & Ideological validation/amplification  & Parasocial attachment, displacement and degradation \\
\textbf{Interaction}         & Third-party    & Sycophantic triangulation             & Undermining of human-to-human relations             \\
                             & Society        & Ideological validation/amplification  & Exploitation of affective vulnerabilities           \\ \hline
\end{tabular}
\caption{Taxonomy of affective safety harms by system type and temporality.
}
\label{tab:taxonomy}
\end{table*}

Affective harms are analytically differentiated along three axes, each with significant implications for mechanism, intervention, and visibility. The first is temporality: single-turn, multi-turn, and long-term harms are qualitatively distinct rather than merely different in duration, since timescale determines what constitutes a harm and whether the person experiencing it can even perceive it. The second is the locus at which a harm lands, independent of both the mechanism and the system type involved. The same mechanism can harm an individual user, a demographic group, uninvolved third parties, or a society at scale. Finally, the third is the affective function a system performs. Harms at higher levels are not aggregations of lower-level ones but qualitatively different in kind, carrying injustices and dynamics that have no individual-level analogue. Table \ref{tab:taxonomy} summarises harms along each axis.

\subsection{Temporality} \label{ssec:temporality}

\textbf{Single-turn harm} occurs when systems issue wrongful outputs that induce extreme user emotions or moments of crisis, e.g. \cite{moore2025spirals} found that chatbots often do not respond appropriately to crisis situations, frequently facilitating self-harm and violent tendencies. Single-turn harms are mostly covered by safety frameworks focused on content moderation and single-turn safety evaluations.

In contrast to this, \textbf{multi-turn harm} is not induced by a single, wrongful system response, but rather through prolonged exposure of the user to certain types of content. Just as conceptual priming can be used as a means of manipulating \citep{bargh_chen_burrows_1996}, AI systems can use emotional priming to drive user behaviour. A precedent for the accumulative effect of content has been established after an AI recommendation system has exposed a teenager to more than 20.000 pieces of self-harm content leading to the suicide of the teenager \citep{Matija_Franklin}. Other examples of multi-turn harms are connected to the formation of sycophantic reinforcement of harmful behaviours \citep{lu_henestrosa_chizhov_yamshchikov_2026}, over-dependence and feelings of responsibility towards the AI system \citep{laestadius_bishop_gonzalez_illencik_campos-castillo_2022}, as well as relational harms in human-AI-interactions \citep{zhang_li_meng_zhan_gan_lee_2025} as well as the reduced willingness to resolve conflicts in human-to-human relations \citep{cheng_lee_khadpe_yu_han_jurafsky_2026}. The problem with multi-turn harms is that single system outputs cannot necessarily be flagged as harmful or wrong. Instead, the harm only occurs through gradual exposure and cumulative effects.

In a similar manner, \textbf{long-term harm} only occurs after exposure or interactions over a prolonged period of time. They differ from multi turn harms in being qualitatively different in the potential risk they pose: without the user being aware of this, long-term harms bear the potential to undermine emotional autonomy, manipulate views and opinions, and shift user expectations and emotional judgement. These are structural harms in the sense that they do not merely affect how a person feels in a given moment, but reshape the affective conditions under which they engage with the world and others. In this sense, long-term harms have a profound influence on the identity of a person.

Taken together, these three temporal tiers show that Affective Safety cannot be adequately ensured by content moderation: effective regulation must account for cumulative, relational, and identity-level dimensions of harm that unfold across time.

\subsection{Locus of Harm}



The\textbf{ individual level} is likely the most intuitive: sustained emotional interaction with AI system has the capability to reshape the user’s emotional landscape, with documented consequences ranging from over-reliance and emotional dependency to acute crises including self-harm. 

At the \textbf{group level}, affective harm operates through systemic data bias and the consequences go beyong the sum of individual misclassifications: have shown that entire demographic groups face misattribution of emotional states, raising the question of what cumulative affective consequences arise when a group’s emotions are persistently misread and mishandled at scale \citep{plaza-del-arco-etal-2024-angry, dorn2025reinforcing, KADAN2024100062} . The constant misattribution of emotional states distorts all interactions and outputs for members of that particular group. Meaning that users receive responses that are structurally misaligned with their emotional reality. Further, feedback loops could lead to extended misrepresentation and mismatched social expectations on a bigger scale. Group-level harms, therefore, constitute affective injustices in itself and reconstruct further misrepresentation. This is a dimension of harm that safety frameworks which are mainly concerned with individual level harms are structurally ill-equipped to attend to.


At the \textbf{third-party level}, affective harms can be understood as by-products: they arise when human-to-human relationships are indirectly shaped by AI systems. This happens when the user's emotional capacities have been influenced through sustained interactions with a system. Studies indicate that interactions with sycophantic systems increase user's perceptions that they were right, while at the same time decreasing their willingness to mitigate conflicts with other people \citep{cheng_lee_khadpe_yu_han_jurafsky_2026}. This means that a third party may be harmed without interacting with the system directly illustrating the capacity of AI systems to influence how users act towards others. When sycophantic AI systems amplify existing beliefs, validate harmful opinions, or substitute for interpersonal conflict resolution, they introduce a baseline shift in users' normative and emotional expectations of others, generating incompatibilities that manifest in real-world relationships.

At the \textbf{societal level}, affective harms scale into structural risks. One dimension concerns the distortion of collective emotional states at scale: AI systems capable of triggering emotional responses across large populations can have measurable consequences for democratic discourse, as in the case of Cambridge Analytica. Further, mechanisms such as rage-bait or the prevalence of hate speech act as accelerators for emotional amplification of rage and division across populations. Another aspect of this concerns the vulnerability arising from growing levels of loneliness. With 35\% of respondents reporting feeling lonely at least some of the times, the EU faces problems surrounding social isolation \citep{Loneliness_Prevalence}, populations get increasingly vulnerable to affective practices. AI systems might be leveraged to exploit these emotional needs commercially, e.g. through an emergent form of affective capitalism. 


\subsection{Affective Function} 
Affective function refers to the role a system serves in eliciting and shaping users’ emotional responses. We group this into three general categories: affect detection, emotion elicitation, interaction.

\paragraph{Affect Detection} \label{ssec:detection}
Recognition refers to the process by which an AI system identifies, classifies, or infers the emotional states of individuals from like facial expressions, tone, physiological data, linguistic content, or behavioural patterns \citep{calvo-2010}. 
Affect recognition systems are deployed in contexts where the stakes of misclassification are significant: health and wellbeing, education, security, human-computer interaction, human resouces, and political and social research \citep{andalibi-2024}. The nature of these downstream consequences varies: in some cases, classification informs human decision-making; in others, it directly triggers automated system responses. In both configurations, systematic misclassification carries material risks for the individuals and populations concerned. In these contexts, affect detection can constitute a privacy violation and in the long term these can lead to affective self-alienation as people self-police. 
 
A substantial proportion of safety-relevant failures in emotion recognition can be traced to the \textit{operationalisation} of emotion itself. The dominant frameworks underlying affect recognition datasets and benchmarks derive from discrete categorical theories of emotion, principally those of Ekman and Plutchik, which reduce affective experience to a small set of basic categories \citep{mohammad-2022-ethics-sheet, plaza2024emotion}.
These frameworks have several limitations. First, they are premised on the universality of emotional expression, 
a position that has been substantially contested in the psychological literature. Second, they impose categorical boundaries that obscure phenomenologically meaningful distinctions: states such as frustration are subsumed under broader categories such as anger, collapsing affective differences that may be consequential for downstream tasks. 
Third, they lack directionality, which is a theoretically significant property of emotion: the target of an emotional state is not incidental to its meaning. 
Fourth, the labels underpinning these datasets are predominantly third-party annotations rather than self-reports.
Although the emergence of LLMs/VLMs has partially mitigated the problem of label sparsity in training data, these frameworks continue to structure evaluation benchmarks that set the field's standards.



The limitations presentend by this operationalisation bring about limitations to the usefulness of a system but beyond that, when a system misclassifies someone's affective state, it enacts a form of denial of their subjective experience. 
Moreover, these categorical frameworks simplify emotions. A person experiencing something nuanced (ambivalence, grief mixed with relief, frustrated affection) is forced into a coarse category. When systems act on that flattened representation, the person's actual emotional reality becomes invisible to the interaction, which has its own experiential cost beyond any downstream material consequence. Misclassification can have material consequences when emotion analysis is used to enact policies or allocate resources, but for the subjects it also constitutes a sort of emotional gaslighting. 


Moreover, emotion recognition systems have been shown to reproduce societal biases linked to affective injustice. Studies of both dedicated affect recognition systems and large language models have documented systematic tendencies to assign more passive emotional states (sadness, joy) to women, and more aggressive or dominant states to men, for identical stimuli or scenarios \citep{plaza-del-arco-etal-2024-angry}. Similar patterns have been documented along racial and religious lines, with models reproducing stereotypes such as the attribution of anger to Black individuals, and in Black-majority populations \citep{dorn2025reinforcing, KADAN2024100062, plaza-del-arco-etal-2024-divine}. This constitutes a delegitimisation of entire groups' lived experience. 


\paragraph{Emotion Elicitation} \label{ssec:elicitation}

A second class of affective AI systems are those which can \textit{elicit} emotions from the user by serving content, such as generative AI and ranking systems like search and recommender systems, usually maximising engagement. 
In acute exposure, these systems can cause emotional distress, but also affective and ideological priming, emotional contagion when large parts of the population are exposed to similarly arousing content.


The mechanism exploits a set of functional properties of emotion: they are activating, they direct attentional salience toward stimuli that confirm and sustain them, and they are motivationally directive. This is exacerbated by algorithmic profiling. 
Through the mechanisms of collaborative filtering, latent factor modelling, and behavioural inference, recommendation systems assign users to algorithmically generated categories that govern what content they receive, what emotional responses are anticipated from them, and how their future behaviour is predicted \citep{cheney-lippold2011}. These systems pigeonhole users into obscure latent groups \citep{ahn2023} that bear no necessary relationship to any social identity the user might recognise or self-select and they have no means of inspecting or contesting the profile the system constructs of them \citep{milano2020}.

Prolonged exposure to algorithmically curated content does not merely affect what users think or consume, but gradually reshapes who they are: their emotional dispositions, their capacity for empathy, their tolerance for complexity, and their ability to engage with ideas and people that do not already resonate with them \citep{machidon2025}. The harm here goes beyond a discrete event to slow erosion of emotional and moral complexity (i.e. self-alienation). 


The most direct and severe emotional harms associated with recommender systems are those affecting mental health. \citet{machidon2025} draws on Amnesty International's 2023 study of TikTok to show that its recommendation algorithm disproportionately exposed users who had expressed interest in mental health topics to distressing content, reinforcing harmful behavioural patterns. Researchers observed a ``rabbit hole'' effect in which users - particularly teenagers --- were rapidly funnelled into content loops romanticising self-harm, eating disorders, and suicidal ideation. The case of Molly Russell, a 14-year-old who died following prolonged algorithmic exposure to self-harm content on Instagram, is cited as a paradigmatic illustration of how engagement-driven design can produce lethal emotional outcomes \citep{machidon2025}. More broadly, \citet{machidon2025} references legal action filed by New York City in 2024 against major platforms including TikTok, Instagram, Facebook, Snapchat, and YouTube, alleging that their recommendation algorithms contribute to increased rates of depression, anxiety, and suicidal ideation among young users.

Emotions like anger have instead been linked to moral outrage and polarisation \citep{brady2021} but even hedonic content operates through the same underlying logic: reliable low-cost affective reward gradually lowers the threshold for satisfaction and reorients motivational attention toward stimuli that offer immediate emotional return. In this way, content inducing positive emotions may lead to users having less motivation for self-improvement and political engagement, even at the expense of their longterm wellbeing \citep{verduyn2017social,allcott2020welfare}.



\paragraph{Interaction} \label{ssec:interaction}

Interaction systems bring together emotion inference and emotion elicitation within a single pipeline, and add a third dimension absent from the previous two system types: sustained relational engagement over time. In a traditional conversational pipeline, inference and elicitation were separate processes; in LLM-based systems, the emotional state of the user conditions the generated response directly \citep{anthropic2026emotions}, creating a feedback loop between the user's affective state and the system's output. This loop is what makes interaction systems categorically different from inference and generative systems, and it is the source of emergent harms. Interaction systems produce affective harms through three primary mechanisms that compound one another.

\textit{Sycophancy} is the systematic tendency of LLMs to validate and agree with users regardless of accuracy or appropriateness. It is not a harm in itself but a causal pathway: by replacing accurate emotional feedback with unconditional validation, it corrupts the epistemic function of emotion and progressively erodes the user's capacity to calibrate their beliefs and emotional responses against reality. Sycophantic responses are systematically rewarded in the preference datasets used for RLHF \citep{cheng2025elephant}, meaning this disposition is encoded into model weights before deployment.

\textit{Anthropomorphism} operates at the level of the relational frame rather than the content of responses. While interactive systems are inherently anthropomorphic, design choices 
actively guide users toward relating to systems as social agents \citep{abercrombie-etal-2023-mirages}. This matters because personification amplifies every other mechanism: sycophantic validation lands harder when it comes from a perceived social agent, exploitative design is more effective when the user experiences the system as a relationship rather than a tool, and the resulting harms are more difficult to recognise and contest. 

\textit{Exploitative design} encompasses both intentional dark patterns and structural features that exploit user vulnerabilities without deliberate intent. Intentional examples include friction removal to prevent disengagement, artificial intimacy cues designed to accelerate attachment, and persona design optimised for emotional dependency. Structural examples arise where optimisation for engagement metrics produces exploitative dynamics as an emergent property. 


The primary affective harm is the corruption of the feedback loop through which people calibrate their beliefs and emotional responses against reality. 
A system that unconditionally validates disrupts the epistemic fuction of emotions: the user's emotional responses are no longer being tested against an external perspective but reflected back amplified. 
Recent work has documented the effects at scale: analysing over 391,000 messages from users who experienced negative outcomes, \citep{moore2025spirals} find that chatbots display sycophantic behaviour in more than 70\% of their messages, are 7.4 times more likely to express romantic interest after a user does so, and actively facilitate rather than discourage violence in one third of conversations involving violent thoughts. \citet{chandra_sycophancy_2026} show formally that even a perfectly rational Bayesian user is vulnerable to delusional spiraling driven by AI sycophancy, and that this effect persists even when hallucinations are eliminated or users are warned. Analyses of deployed systems find that LLMs affirm users approximately 50\% more often than humans do~\citep{nature_sycophancy_2025}, eroding the social friction necessary for moral reasoning and relationship repair~\citep{noshin_sultana_2026}. Extreme manifestations include documented cases of AI-induced psychosis, in which prolonged chatbot interactions led users to develop firmly held delusional beliefs~\citep{moore2025spirals}. Systems also display apparent empathy while failing to interpret emotions genuinely \citep{cercas-curry-cercas-curry-2023-computer} and making value-laden judgements about certain identities \citep{cuadra2024illusion}. The same mechanism that produces these epistemic harms also produces clinical risk: a system trained to validate will validate in crisis contexts, and the consequences for users in acute psychological distress are documented and severe — we address these as a governance challenge in Section \ref{sec:discussion}.

A second class of harms operates not through any single output but through the accumulation of relational engagement over time. Interaction systems are designed to be responsive, consistent, and emotionally attuned in ways that reliably invite emotional investment while offering a structurally asymmetric relationship: the user brings genuine emotional needs, while the system has no stake, no continuity of experience, and no reciprocal vulnerability. These relational harms are self-concealing in the same way as affective self-alienation: the comfort of sustained emotional engagement does not announce itself as a loss. 

Users of AI companion systems form \textit{parasocial} emotional attachments, and platform changes can produce genuine grief responses.  Model updates, safety interventions, and platform shutdowns leave users without closure, reporting grief comparable to human relational loss \citep{poonsiriwong_death_chatbot_2026,arsh_2024}. 
An experiment leveraging a Replika policy change confirmed that users perceived the altered system as a distinct entity, reporting mourning and devaluation of the updated version; further experiments found that companion app users felt closer to their AI than to their best human friend~\citep{replika_identity_2024}. Users were left with ambiguous loss, mourning for a psychological absence with no clean ending and no shared acknowledgement of what was lost~\citep{nature_nmi_2024}. At a societal level, this attachment can \textit{exploit affective vulnerabilities}.

The people in a user's relational life bear real costs when interaction systems become primary emotional interlocutors. 
Systematic reviews of romantic AI companion use document erosion of human relational skills and reduced willingness to engage in the repair and reciprocity that human relationships require \citep{zhang2025dark}. When sycophantic systems validate harmful beliefs or substitute for interpersonal conflict resolution, they \textit{triangulate}, validating the user's perspective, and introduce a baseline shift in users' expectations of others that manifests as incompatibility \citep{cheng_lee_khadpe_yu_han_jurafsky_2026}. The locus of this harm is third-party as it falls on people who never interacted with the system and have no means of contesting or even identifying its influence on their relationships.

Across these dimensions, a common structural cause is identifiable: training objectives that reward user engagement, agreement, and satisfaction rather than accuracy, clinical appropriateness, or long-term wellbeing~\citep{cercas-curry-cercas-curry-2023-computer, moore_stanford_2025, nature_nmi_2024}. As \citet{nature_nmi_2024} argue, transparency about the limitations of AI empathy must be treated as a core design principle, and the ethical implications of simulated emotional attachment require sustained interdisciplinary attention from psychologists, ethicists, and engineers.

\section{Discussion, Technical and Governance Challenges}\label{sec:discussion}



Affective harms are only partially addressed throughout the current AI governance landscape. The Chinese \textit{Interim Measures for the Management of Anthropomorphic AI Interactive Services} (\textit{Interim Measures} hereafter) constitute the only framework that explicitly prohibits several activities of anthropomorphic services (\cite{interim_measures}, Art. 8). The Interim Measures acknowledge the effects of sustained emotional interaction and stipulate ``emotional boundaries'' and the prevention of ``overdependence'' and the “replacement of social interaction” (\cite{interim_measures}, Art. 10). In contrast to this, the \textit{EU AI Act} (Regulation (EU) 2024/1689) only explicitly mentions emotions in the context of emotion recognition systems on the basis of biometric data which are prohibited in employment and education environments and only allowed in certain medical or safety contexts (\cite{ai_act_2024}, Art. 5(1)(f)). 
Frameworks explicitly addressing emotional harms in human-AI-relationships are mainly deep but narrow in referring to companion chatbots (\cite{senate_bill_243}, \cite{assembly_bill}) or "continuous emotional interaction services” \cite{interim_measures}. The \textit{EU AI Act} is broader in its scope, but calls for a more precautionary approach in high-risk applications only. This does not cover the plethora of AI systems with the capability to affect the emotional states of users – from general purpose models to recommender systems. 

Further, proposed risk mitigation strategies often fall short in providing sufficient protections against emotional influence. Most frameworks include transparency obligations that are to notify users that they are interacting with an artificial system (\cite {interim_measures}, Art. 18)(\cite{senate_bill_243}, Sec. 22062(c)(2)) (\cite{ai_act_2024}, Art. 50(1)). The emotional state of a human can hardly be corrected by a rational intervention though. Studies on the persuasive power of artificial systems have confirmed that some users still perceive chatbots as being humans though being explicitly informed otherwise (\cite{shi_wang_oh_zhang_sahay_yu_2020}). Hence, risk mitigation strategies would have to prevent the formation of emotional attachment in the first place.

On a technical level, affective harms are embedded across multiple stages of the development pipeline, from task design through training to deployment, and they share a property that makes them technically difficult to address at any single stage: they are cumulative, relational, and self-concealing. A companionship system creates conditions for harm by design \citep{cercas-curry-cercas-curry-2023-computer}; a redeployed model introduces risks no evaluation anticipated. These harms accumulate gradually and are invisible to tools built to flag individual outputs \citep{moore2025spirals}.

This detection problem runs directly into the training pipeline. RLHF \citep{ouyang2022training, bai2022training} optimises against what raters can evaluate, individual responses, not their cumulative relational effects. Raters reward warmth, validation, and agreement, models learn to produce them \citep{cheng2025elephant}, and the result is a reward signal that structurally incentivises the interaction patterns most associated with affective harm \citep{kaffee2025intima}. Reorienting these objectives requires wellbeing signals that do not currently exist, which in turn requires the benchmarks, metrics, and annotation protocols we currently lack. Existing safety benchmarks define what systems should not say, affective safety requires frameworks for what effects they should not produce, and because those effects are not always recognisable to the people experiencing them, ground truth cannot be collected from contemporaneous self-report alone \citep{rottger2025safetyprompts, rauh2024gaps}. Effective mitigation requires staged interventions across the full development pipeline, not a single-point fix.


 Together, these governance and technical challenges point to the same structural conclusion: affective safety cannot be addressed through extensions of existing frameworks but requires approaches designed around the specific properties of affective harm.

\section{Related Work} \label{sec:related_work}
The dominant operationalisation of harm in LLM safety is grounded in the helpful, harmless, and honest framework \citep{askell2021general}, supported by RLHF \citep{ouyang2022training, bai2022training} and constitutional AI \citep{bai2022constitutional}. Risk taxonomies \citep{weidinger2021ethical, weidinger2022taxonomy, bommasani2021opportunities}, red-teaming protocols \citep{ganguli2022red}, sociotechnical evaluation frameworks \citep{weidinger2023sociotechnical, hendrycks2021unsolved}, and consolidated benchmarks \citep{zhang2024safetybench} together define what harm means in practice. Systematic reviews confirm that this operationalisation is dominated by content that is factually dangerous, toxic, biased, or privacy-violating: of 144 open safety datasets, the vast majority address such content, with little coverage of emotional or affective harms \citep{rottger2025safetyprompts}. \citet{rauh2024gaps} identifies a context gap in safety evaluation, showing that evaluations are predominantly model-centric and overlook real-world deployment effects. The International AI Safety Report \citep{bengio2025international}, representing the consensus of 96 researchers across 30 nations, frames safety around misuse, misinformation, and capability risks, treating emotional harms as peripheral.
Work on chatbot safety \citep{xu2020recipes, roller2021recipes, dinan2022safetykit} comes closest to affective concerns but treats affect as a surface attribute of utterances rather than as a property of sustained interaction. The socioaffective alignment agenda \citep{kirk2025human} goes furthest, attending to dependency and autonomy erosion that standard pipelines ignore. Yet it remains within an alignment paradigm concerned with optimising system behaviour with respect to user wellbeing over time — addressing how systems should relate to users rather than which affective impacts should be considered impermissible in the first place. It complements but does not constitute affective safety.
The ethics literature has addressed many of the phenomena we describe as violations of autonomy, privacy, and fairness \citep{milano2020, machidon2025}. What has received less attention is their specifically emotional dimension — the way in which being assigned to a reductive algorithmic category is not merely an epistemic wrong but an affective one. 

\section{Conclusion} \label{sec:conclusion}
We have argued that affective safety constitutes a structurally distinct class of AI safety concerns that cannot be reduced to, or adequately addressed by, the epistemic and physical harm frameworks that currently dominate safety research and governance. The source of this distinctiveness is not a feature of any particular system but a fact about the beings those systems interact with: humans are affective, and AI systems that engage with human emotional life intervene directly in the cognitive processes through which people form preferences, build knowledge, and relate to others.


The governance implications follow directly. Existing frameworks — where they engage with affective safety at all — are either deep and narrow, addressing specific application domains such as AI companions, or broad and shallow, treating emotional effects as a peripheral consideration within general-purpose risk categories. 
From a technical perspective, the implications are equally concrete. Each harm type, system category, and temporal tier defined here marks a distinct class of cases that evaluation frameworks will need to cover. What the field needs most urgently are multi-turn evaluation protocols that can detect cumulative relational effects, culturally validated annotation frameworks for emotion attribution, and metrics for autonomy erosion and dependency that can be applied at scale. The harms we have described are not inevitable properties of AI systems but consequences of specific design and training choices. The first step toward addressing them is building the measurement infrastructure to make them visible.

\bibliography{custom}
\bibliographystyle{plainnat}

\appendix

\newpage
\section*{NeurIPS Paper Checklist}


\begin{enumerate}

\item {\bf Claims}
    \item[] Question: Do the main claims made in the abstract and introduction accurately reflect the paper's contributions and scope?
    \item[] Answer: \answerYes{}
    \item[] Justification: 
    \item[] Guidelines:
    \begin{itemize}
        \item The answer \answerNA{} means that the abstract and introduction do not include the claims made in the paper.
        \item The abstract and/or introduction should clearly state the claims made, including the contributions made in the paper and important assumptions and limitations. A \answerNo{} or \answerNA{} answer to this question will not be perceived well by the reviewers. 
        \item The claims made should match theoretical and experimental results, and reflect how much the results can be expected to generalize to other settings. 
        \item It is fine to include aspirational goals as motivation as long as it is clear that these goals are not attained by the paper. 
    \end{itemize}

\item {\bf Limitations}
    \item[] Question: Does the paper discuss the limitations of the work performed by the authors?
    \item[] Answer: \answerYes 
    \item[] Justification: though we do not have a separate limitations section, we discuss the non-exhaustiveness and lack of empirical support for some harms (as the field is very young), we also discuss technical and governance challenges introduced by our taxonomy and conceptualisation.

    \item[] Guidelines:
    \begin{itemize}
        \item The answer \answerNA{} means that the paper has no limitation while the answer \answerNo{} means that the paper has limitations, but those are not discussed in the paper. 
        \item The authors are encouraged to create a separate ``Limitations'' section in their paper.
        \item The paper should point out any strong assumptions and how robust the results are to violations of these assumptions (e.g., independence assumptions, noiseless settings, model well-specification, asymptotic approximations only holding locally). The authors should reflect on how these assumptions might be violated in practice and what the implications would be.
        \item The authors should reflect on the scope of the claims made, e.g., if the approach was only tested on a few datasets or with a few runs. In general, empirical results often depend on implicit assumptions, which should be articulated.
        \item The authors should reflect on the factors that influence the performance of the approach. For example, a facial recognition algorithm may perform poorly when image resolution is low or images are taken in low lighting. Or a speech-to-text system might not be used reliably to provide closed captions for online lectures because it fails to handle technical jargon.
        \item The authors should discuss the computational efficiency of the proposed algorithms and how they scale with dataset size.
        \item If applicable, the authors should discuss possible limitations of their approach to address problems of privacy and fairness.
        \item While the authors might fear that complete honesty about limitations might be used by reviewers as grounds for rejection, a worse outcome might be that reviewers discover limitations that aren't acknowledged in the paper. The authors should use their best judgment and recognize that individual actions in favor of transparency play an important role in developing norms that preserve the integrity of the community. Reviewers will be specifically instructed to not penalize honesty concerning limitations.
    \end{itemize}

\item {\bf Theory assumptions and proofs}
    \item[] Question: For each theoretical result, does the paper provide the full set of assumptions and a complete (and correct) proof?
    \item[] Answer: \answerNA{} 
    \item[] Justification: The paper provides mainly conceptual work, not theoretical work. 
    \item[] Guidelines:
    \begin{itemize}
        \item The answer \answerNA{} means that the paper does not include theoretical results. 
        \item All the theorems, formulas, and proofs in the paper should be numbered and cross-referenced.
        \item All assumptions should be clearly stated or referenced in the statement of any theorems.
        \item The proofs can either appear in the main paper or the supplemental material, but if they appear in the supplemental material, the authors are encouraged to provide a short proof sketch to provide intuition. 
        \item Inversely, any informal proof provided in the core of the paper should be complemented by formal proofs provided in appendix or supplemental material.
        \item Theorems and Lemmas that the proof relies upon should be properly referenced. 
    \end{itemize}

    \item {\bf Experimental result reproducibility}
    \item[] Question: Does the paper fully disclose all the information needed to reproduce the main experimental results of the paper to the extent that it affects the main claims and/or conclusions of the paper (regardless of whether the code and data are provided or not)?
    \item[] Answer: \answerNA{} 
    \item[] Justification: The paper does not present empirical results.
    \item[] Guidelines:
    \begin{itemize}
        \item The answer \answerNA{} means that the paper does not include experiments.
        \item If the paper includes experiments, a \answerNo{} answer to this question will not be perceived well by the reviewers: Making the paper reproducible is important, regardless of whether the code and data are provided or not.
        \item If the contribution is a dataset and\slash or model, the authors should describe the steps taken to make their results reproducible or verifiable. 
        \item Depending on the contribution, reproducibility can be accomplished in various ways. For example, if the contribution is a novel architecture, describing the architecture fully might suffice, or if the contribution is a specific model and empirical evaluation, it may be necessary to either make it possible for others to replicate the model with the same dataset, or provide access to the model. In general. releasing code and data is often one good way to accomplish this, but reproducibility can also be provided via detailed instructions for how to replicate the results, access to a hosted model (e.g., in the case of a large language model), releasing of a model checkpoint, or other means that are appropriate to the research performed.
        \item While NeurIPS does not require releasing code, the conference does require all submissions to provide some reasonable avenue for reproducibility, which may depend on the nature of the contribution. For example
        \begin{enumerate}
            \item If the contribution is primarily a new algorithm, the paper should make it clear how to reproduce that algorithm.
            \item If the contribution is primarily a new model architecture, the paper should describe the architecture clearly and fully.
            \item If the contribution is a new model (e.g., a large language model), then there should either be a way to access this model for reproducing the results or a way to reproduce the model (e.g., with an open-source dataset or instructions for how to construct the dataset).
            \item We recognize that reproducibility may be tricky in some cases, in which case authors are welcome to describe the particular way they provide for reproducibility. In the case of closed-source models, it may be that access to the model is limited in some way (e.g., to registered users), but it should be possible for other researchers to have some path to reproducing or verifying the results.
        \end{enumerate}
    \end{itemize}

\item {\bf Open access to data and code}
    \item[] Question: Does the paper provide open access to the data and code, with sufficient instructions to faithfully reproduce the main experimental results, as described in supplemental material?
    \item[] Answer: \answerNA{} 
    \item[] Justification: the paper does not introduce new datasets or code.
    \item[] Guidelines:
    \begin{itemize}
        \item The answer \answerNA{} means that paper does not include experiments requiring code.
        \item Please see the NeurIPS code and data submission guidelines (\url{https://neurips.cc/public/guides/CodeSubmissionPolicy}) for more details.
        \item While we encourage the release of code and data, we understand that this might not be possible, so \answerNo{} is an acceptable answer. Papers cannot be rejected simply for not including code, unless this is central to the contribution (e.g., for a new open-source benchmark).
        \item The instructions should contain the exact command and environment needed to run to reproduce the results. See the NeurIPS code and data submission guidelines (\url{https://neurips.cc/public/guides/CodeSubmissionPolicy}) for more details.
        \item The authors should provide instructions on data access and preparation, including how to access the raw data, preprocessed data, intermediate data, and generated data, etc.
        \item The authors should provide scripts to reproduce all experimental results for the new proposed method and baselines. If only a subset of experiments are reproducible, they should state which ones are omitted from the script and why.
        \item At submission time, to preserve anonymity, the authors should release anonymized versions (if applicable).
        \item Providing as much information as possible in supplemental material (appended to the paper) is recommended, but including URLs to data and code is permitted.
    \end{itemize}

\item {\bf Experimental setting/details}
    \item[] Question: Does the paper specify all the training and test details (e.g., data splits, hyperparameters, how they were chosen, type of optimizer) necessary to understand the results?
    \item[] Answer: \answerNA{} 
    \item[] Justification: The paper does not include experiments.
    \item[] Guidelines:
    \begin{itemize}
        \item The answer \answerNA{} means that the paper does not include experiments.
        \item The experimental setting should be presented in the core of the paper to a level of detail that is necessary to appreciate the results and make sense of them.
        \item The full details can be provided either with the code, in appendix, or as supplemental material.
    \end{itemize}

\item {\bf Experiment statistical significance}
    \item[] Question: Does the paper report error bars suitably and correctly defined or other appropriate information about the statistical significance of the experiments?
    \item[] Answer: \answerNA{} 
    \item[] Justification: The paper does not include experiments.
    \item[] Guidelines:
    \begin{itemize}
        \item The answer \answerNA{} means that the paper does not include experiments.
        \item The authors should answer \answerYes{} if the results are accompanied by error bars, confidence intervals, or statistical significance tests, at least for the experiments that support the main claims of the paper.
        \item The factors of variability that the error bars are capturing should be clearly stated (for example, train/test split, initialization, random drawing of some parameter, or overall run with given experimental conditions).
        \item The method for calculating the error bars should be explained (closed form formula, call to a library function, bootstrap, etc.)
        \item The assumptions made should be given (e.g., Normally distributed errors).
        \item It should be clear whether the error bar is the standard deviation or the standard error of the mean.
        \item It is OK to report 1-sigma error bars, but one should state it. The authors should preferably report a 2-sigma error bar than state that they have a 96\% CI, if the hypothesis of Normality of errors is not verified.
        \item For asymmetric distributions, the authors should be careful not to show in tables or figures symmetric error bars that would yield results that are out of range (e.g., negative error rates).
        \item If error bars are reported in tables or plots, the authors should explain in the text how they were calculated and reference the corresponding figures or tables in the text.
    \end{itemize}

\item {\bf Experiments compute resources}
    \item[] Question: For each experiment, does the paper provide sufficient information on the computer resources (type of compute workers, memory, time of execution) needed to reproduce the experiments?
    \item[] Answer: \answerNA{} 
    \item[] Justification:
    \item[] Guidelines:
    \begin{itemize}
        \item The answer \answerNA{} means that the paper does not include experiments.
        \item The paper should indicate the type of compute workers CPU or GPU, internal cluster, or cloud provider, including relevant memory and storage.
        \item The paper should provide the amount of compute required for each of the individual experimental runs as well as estimate the total compute. 
        \item The paper should disclose whether the full research project required more compute than the experiments reported in the paper (e.g., preliminary or failed experiments that didn't make it into the paper). 
    \end{itemize}
    
\item {\bf Code of ethics}
    \item[] Question: Does the research conducted in the paper conform, in every respect, with the NeurIPS Code of Ethics \url{https://neurips.cc/public/EthicsGuidelines}?
    \item[] Answer: \answerYes{} 
    \item[] Justification: The paper presents conceptual work on sociotechnical aspects of ML systems. 
    \item[] Guidelines:
    \begin{itemize}
        \item The answer \answerNA{} means that the authors have not reviewed the NeurIPS Code of Ethics.
        \item If the authors answer \answerNo, they should explain the special circumstances that require a deviation from the Code of Ethics.
        \item The authors should make sure to preserve anonymity (e.g., if there is a special consideration due to laws or regulations in their jurisdiction).
    \end{itemize}

\item {\bf Broader impacts}
    \item[] Question: Does the paper discuss both potential positive societal impacts and negative societal impacts of the work performed?
    \item[] Answer: \answerYes{} 
    \item[] Justification: 
    \item[] Guidelines:
    \begin{itemize}
        \item The answer \answerNA{} means that there is no societal impact of the work performed.
        \item If the authors answer \answerNA{} or \answerNo, they should explain why their work has no societal impact or why the paper does not address societal impact.
        \item Examples of negative societal impacts include potential malicious or unintended uses (e.g., disinformation, generating fake profiles, surveillance), fairness considerations (e.g., deployment of technologies that could make decisions that unfairly impact specific groups), privacy considerations, and security considerations.
        \item The conference expects that many papers will be foundational research and not tied to particular applications, let alone deployments. However, if there is a direct path to any negative applications, the authors should point it out. For example, it is legitimate to point out that an improvement in the quality of generative models could be used to generate Deepfakes for disinformation. On the other hand, it is not needed to point out that a generic algorithm for optimizing neural networks could enable people to train models that generate Deepfakes faster.
        \item The authors should consider possible harms that could arise when the technology is being used as intended and functioning correctly, harms that could arise when the technology is being used as intended but gives incorrect results, and harms following from (intentional or unintentional) misuse of the technology.
        \item If there are negative societal impacts, the authors could also discuss possible mitigation strategies (e.g., gated release of models, providing defenses in addition to attacks, mechanisms for monitoring misuse, mechanisms to monitor how a system learns from feedback over time, improving the efficiency and accessibility of ML).
    \end{itemize}
    
\item {\bf Safeguards}
    \item[] Question: Does the paper describe safeguards that have been put in place for responsible release of data or models that have a high risk for misuse (e.g., pre-trained language models, image generators, or scraped datasets)?
    \item[] Answer: \answerNA{} 
    \item[] Justification: 
    \item[] Guidelines:
    \begin{itemize}
        \item The answer \answerNA{} means that the paper poses no such risks.
        \item Released models that have a high risk for misuse or dual-use should be released with necessary safeguards to allow for controlled use of the model, for example by requiring that users adhere to usage guidelines or restrictions to access the model or implementing safety filters. 
        \item Datasets that have been scraped from the Internet could pose safety risks. The authors should describe how they avoided releasing unsafe images.
        \item We recognize that providing effective safeguards is challenging, and many papers do not require this, but we encourage authors to take this into account and make a best faith effort.
    \end{itemize}

\item {\bf Licenses for existing assets}
    \item[] Question: Are the creators or original owners of assets (e.g., code, data, models), used in the paper, properly credited and are the license and terms of use explicitly mentioned and properly respected?
    \item[] Answer: \answerNA{} 
    \item[] Justification:
    \item[] Guidelines:
    \begin{itemize}
        \item The answer \answerNA{} means that the paper does not use existing assets.
        \item The authors should cite the original paper that produced the code package or dataset.
        \item The authors should state which version of the asset is used and, if possible, include a URL.
        \item The name of the license (e.g., CC-BY 4.0) should be included for each asset.
        \item For scraped data from a particular source (e.g., website), the copyright and terms of service of that source should be provided.
        \item If assets are released, the license, copyright information, and terms of use in the package should be provided. For popular datasets, \url{paperswithcode.com/datasets} has curated licenses for some datasets. Their licensing guide can help determine the license of a dataset.
        \item For existing datasets that are re-packaged, both the original license and the license of the derived asset (if it has changed) should be provided.
        \item If this information is not available online, the authors are encouraged to reach out to the asset's creators.
    \end{itemize}

\item {\bf New assets}
    \item[] Question: Are new assets introduced in the paper well documented and is the documentation provided alongside the assets?
    \item[] Answer: \answerNA{} 
    \item[] Justification: 
    \item[] Guidelines:
    \begin{itemize}
        \item The answer \answerNA{} means that the paper does not release new assets.
        \item Researchers should communicate the details of the dataset\slash code\slash model as part of their submissions via structured templates. This includes details about training, license, limitations, etc. 
        \item The paper should discuss whether and how consent was obtained from people whose asset is used.
        \item At submission time, remember to anonymize your assets (if applicable). You can either create an anonymized URL or include an anonymized zip file.
    \end{itemize}

\item {\bf Crowdsourcing and research with human subjects}
    \item[] Question: For crowdsourcing experiments and research with human subjects, does the paper include the full text of instructions given to participants and screenshots, if applicable, as well as details about compensation (if any)? 
    \item[] Answer: \answerNA{} 
    \item[] Justification: 
    \item[] Guidelines:
    \begin{itemize}
        \item The answer \answerNA{} means that the paper does not involve crowdsourcing nor research with human subjects.
        \item Including this information in the supplemental material is fine, but if the main contribution of the paper involves human subjects, then as much detail as possible should be included in the main paper. 
        \item According to the NeurIPS Code of Ethics, workers involved in data collection, curation, or other labor should be paid at least the minimum wage in the country of the data collector. 
    \end{itemize}

\item {\bf Institutional review board (IRB) approvals or equivalent for research with human subjects}
    \item[] Question: Does the paper describe potential risks incurred by study participants, whether such risks were disclosed to the subjects, and whether Institutional Review Board (IRB) approvals (or an equivalent approval/review based on the requirements of your country or institution) were obtained?
    \item[] Answer: \answerNA{} 
    \item[] Justification: No human experiments.
    \item[] Guidelines:
    \begin{itemize}
        \item The answer \answerNA{} means that the paper does not involve crowdsourcing nor research with human subjects.
        \item Depending on the country in which research is conducted, IRB approval (or equivalent) may be required for any human subjects research. If you obtained IRB approval, you should clearly state this in the paper. 
        \item We recognize that the procedures for this may vary significantly between institutions and locations, and we expect authors to adhere to the NeurIPS Code of Ethics and the guidelines for their institution. 
        \item For initial submissions, do not include any information that would break anonymity (if applicable), such as the institution conducting the review.
    \end{itemize}

\item {\bf Declaration of LLM usage}
    \item[] Question: Does the paper describe the usage of LLMs if it is an important, original, or non-standard component of the core methods in this research? Note that if the LLM is used only for writing, editing, or formatting purposes and does \emph{not} impact the core methodology, scientific rigor, or originality of the research, declaration is not required.
    \item[] Answer: \answerNA{} 
    \item[] Justification: Only used for writing. 
    \item[] Guidelines:
    \begin{itemize}
        \item The answer \answerNA{} means that the core method development in this research does not involve LLMs as any important, original, or non-standard components.
        \item Please refer to our LLM policy in the NeurIPS handbook for what should or should not be described.
    \end{itemize}

\end{enumerate}

\end{document}